\documentclass[final,3p,times]{elsarticle}

\usepackage{framed,multirow}

\usepackage{amssymb}
\usepackage{amsmath}
\usepackage{latexsym}

\usepackage{url}
\usepackage[dvipsnames]{xcolor}
\definecolor{newcolor}{rgb}{.8,.349,.1}
\usepackage[utf8]{inputenx}
\usepackage{graphicx}
\usepackage{amsmath,amssymb,mathrsfs}
\usepackage{fancyhdr}
\usepackage{calc}
\usepackage[T1]{fontenc}
\usepackage{mathptmx}
\usepackage{physics}
\usepackage{lineno}
\usepackage{tikz}
\usepackage{pgfplots}

\def\ww{\kappa}

\def\nonu{\nonumber}
\def\normU{|| \overline{U} ||}

\newcommand{\bes}{\begin{equation*}}
\newcommand{\ees}{\end{equation*}}
\newcommand{\beq}{\begin{equation}}
\newcommand{\eeq}{\end{equation}}
\newcommand{\bea}{\begin{eqnarray}}
\newcommand{\eea}{\end{eqnarray}}
\newcommand{\beas}{\begin{eqnarray*}}
\newcommand{\eeas}{\end{eqnarray*}}

\newcommand{\gam}{\gamma}
\newcommand{\gamo}{\gamma_1}
\newcommand{\bfA}{{\bf A}}
\newcommand{\bfB}{{\bf B}}
\newcommand{\bfI}{{\bf I}}
\newcommand{\bfK}{{\bf K}}
\newcommand{\bzeta}{\pmb{\zeta}}
\newcommand{\colb}[1]{\textcolor{black}{#1}}
\newcommand{\colt}[1]{\textcolor{black}{#1}}
\def\cp{{\itshape{C}$^+$}}
\def\cm{{\itshape{C}$^-$}}
\def\st{{\itshape{S}}}

\begin{document}

\begin{frontmatter}

\title{Adjoint and direct characteristic equations for two-dimensional compressible Euler flows.}

\address[rvt1]{D\'epartement a\'erodynamique, a\'ero\'elasticit\'e et acoustique (DAAA), ONERA, Université Paris Saclay,
  29 avenue de la Division Leclerc, Ch\^atillon, 92322, France}
\address[rvt3]{D\'epartement a\'erodynamique, a\'ero\'elasticit\'e et acoustique (DAAA), ONERA, Université Paris Saclay,
  8 Rue des Vertugadins, Meudon, 92190, France}
\author[rvt1]{Kevin Ancourt}
\ead{kevin.ancourt@onera.fr}

\author[rvt1]{Jacques Peter\corref{cor1}}
\ead{jacques.peter@onera.fr}
\cortext[cor1]{Corresponding author. Tel.: +33 1 46 73 41 84.}

\author[rvt3]{Olivier Atinault}
\ead{olivier.atinault@onera.fr}

\begin{abstract}
The method of characteristics is a classical method for gaining understanding
  in the solution of a partial differential equation. It has recently been applied
  to the adjoint
  equations of the 2D Euler equations and the first goal of this paper is to present
  a linear algebra
  analysis that greatly simplifies the discussion of the number of independant characteristic
  equations satisfied along a family of characteristic curves. This method may
  be applied for both the direct and the adjoint problem and our second goal is
  to directly derive in conservative variables
  the characteristic equations of 2D compressible inviscid flows.
  Finally, the theoretical results are assessed for a nozzle flow with a
  classical scheme and its dual consistent discrete adjoint.
\end{abstract}   


\end{frontmatter}

\section{Introduction}
The method of characteristics is a well-known method for studying partial differential equations
(PDE). It aims to exhibit specific hypersurfaces in the input domain where the solution of
the PDE of interest satisfies an ordinary differential equation (ODE). When applied to 2D
inviscid compressible flows, it is known to provide a full resolution of a supersonic area only based on the
knowledge of the inflow (whereas it provides partial information for a subsonic
flow) \cite{Fer_46,Sha_54,BonLun_89,Del_08}.
Both theoretical understanding and pratical calculations of a variety of flows
(in nozzles, along steps, along curved walls) are enabled by this technique.
\\
Besides, discrete and continuous adjoint are now well-etablished methods for shape optimization
 \cite{Jam_88,BreGau_04,BreDwi_05,JPDwi_10,SchIliSch_11} and
 goal-oriented mesh adaptation \cite{VenDar_02,Dwi_08,TodVonBou_16}. 
 Because of these important applications, regular efforts have been devoted to the fast and safe writing
 of adjoint modules \cite{GilDutMul_01,GilDutMul_03,EcoAloAlb_17,SagAlbGau_19} and to
 the efficient solving of the adjoint equations \cite{PinMon_13,XuRadMey_15}.
 Adjoint methods are also useful for
 flow control \cite{Lio_68,SarMetSip_15}, meta-modelling \cite{Lau_08,BaaSchDwi_15},
 receptivity-sensitivity-stability analyses \cite{SipMarMel_10,LucBot_14},
 data assimilation \cite{ParDur_15,TalCou_87}.
 In a recent paper \cite{JPDes_22} the characteristic equations (CE) for the adjoint 2D
 Euler equations have been derived.
 \\
 The classical direct characteristic equations (DCE) for 2D inviscid compressible flows
 and the recently presented corresponding ajoint characteristic equations (ACE)
 appear to be linked; In particular,
 the characteristic curves are the same for both problems \cite{JPDes_22}. The analogies between
 (ACE) and (DCE) are further studied in section 2
 where several properties of a generic sytem of equations embbeding (ACE) and (DCE) are demonstrated.
 In particular, this formal linear algebra approach greatly simplifies the determination of the number
 of independant (CE) satisfied along two families of characteristic curves.
 \\
 Whereas the (ACE) need to be derived from the complete 8x8 linear system for the
 Cauchy problem -- the (DCE) have been derived with various mechanical asumptions and also
 various sets of variables.
 For the sake of a full understanding, the direct characteristic problem is solved in conservative
 variables from the complete
 Cauchy problem and the resulting differential forms are discussed in section  3. 
 \\
 Finally, an inviscid supersonic nozzle flow is considered in section 4 where both the (DCE) and
 the (ACE) are numerically assessed. 
 %
%
\section{Common properties of the linear systems for the Cauchy direct and adjoint problems}
%
%
%
\subsection{Notations}
%
%
 We denote by $W = (\rho, \rho u , \rho v , \rho E)$ the conservative variables,  with $\rho$ the density, $(u,v)$ the components
  of the velocity $\overline{U}$, by $e$ the internal energy, by $E$ the total energy.
A thermally and calorically perfect gas law is considered. The static pressure $p$, total enthalpy $H$
 and entropy $S$ read 
\begin{linenomath}
$$
p = (\gamma - 1)\rho e = (\gamma - 1)(\rho E - 0.5  \rho \normU ^2)  
$$
$$
H = E + \frac{p}{\rho}  \qquad \qquad S = c_v ln(\frac{p}{\rho^{\gamma}})
$$
\end{linenomath}
with a constant $\gamma = 1.4$ and a constant heat capacity at constant volume, $c_v$.
A subscript $i$ is used to denote the classical stagnation quantities $(\rho_i,p_i,T_i)$.
Finally, the adjoint vector of the functional output of interest is denoted as $\psi =(\psi_1, \psi_2,\psi_3,\psi_4 )$.
 %
 %
\subsection{Generic Cauchy problem embbeding (ACE) and (DCE)}
%
%
Given $a$ and $b$, two neighboring points in the fluid domain,
we denote by $(dx,dy)=\overrightarrow{ab}$.  The increment in the adjoint components
 between the two points are denoted as $( d\psi_1 ,  d\psi_2 ,  d\psi_3 ,  d\psi_4 ) =
(\psi_1^b - \psi_1^a,  \psi_2^b - \psi_2^a,  \psi_3^b - \psi_3^a, \psi_4^b - \psi_4^a)$ . The Cauchy problem for the adjoint vector
aims at calculating its derivatives at point $a$ with first order in space, and the singularities of this problem
 are at the core of the search of the (CE).
 The corresponding 8$\times$8 linear system associates first-order Taylor expansions
 and the 2D adjoint Euler equations.
 It reads 
\begin{linenomath}
\beq
\begin{bmatrix}
dx & 0  & 0  & 0  & dy & 0  & 0  & 0 \\
0  & dx & 0  & 0  & 0  & dy & 0  & 0 \\
0  & 0  & dx & 0  & 0  & 0  & dy & 0 \\
0  & 0  & 0  & dx & 0  & 0  & 0  & dy \\
   &    &    &    &    &    &    &   \\
   &    &    &    &    &    &    &   \\
   & -  & A^T  &    &    & -  & B^T  &   \\
   &    &    &    &    &    &    &   
\end{bmatrix}
\begin{bmatrix}
(\partial \psi_1/\partial x) \\
(\partial \psi_2/\partial x) \\
(\partial \psi_3/\partial x) \\
(\partial \psi_4/\partial x) \\
(\partial \psi_1/\partial y) \\
(\partial \psi_2/\partial y) \\
(\partial \psi_3/\partial y) \\
(\partial \psi_4/\partial y) 
\end{bmatrix}
=
\begin{bmatrix}
 d\psi_1 \\
 d\psi_2 \\
 d\psi_3 \\
 d\psi_4 \\
 0 \\
 0 \\
 0\\
 0
\end{bmatrix}
\label{e_cadj}
\eeq
\end{linenomath}
where $-A^T$ and $-B^T$ are the opposite and transposed Jacobians of the Euler fluxes.
The straighforward counterpart for the derivatives of the discrete flow field $W$,
appears in equation (\ref{e_cbase}).
Several results regarding the minors of the matrices in (\ref{e_cadj}) and (\ref{e_cbase}) 
 and also the number of independant (CE) when ($dx$,$dy$) make the 8$\times$8 matrices singular
 may by demonstrated considering a generic matrix
\begin{linenomath}
$$
\bfK = \begin{bmatrix}
dx & 0  & 0  & 0  & dy & 0  & 0  & 0 \\
0  & dx & 0  & 0  & 0  & dy & 0  & 0 \\
0  & 0  & dx & 0  & 0  & 0  & dy & 0 \\
0  & 0  & 0  & dx & 0  & 0  & 0  & dy \\
   &    &    &    &    &    &    &   \\
   &    &    &    &    &    &    &   \\
   &   & \bfA  &    &    &   & \bfB  &   \\
   &    &    &    &    &    &    &   
\end{bmatrix}
$$
\end{linenomath}
where $({\bf A}, {\bf B})$ stands either for $(-A^T, -B^T)$ or for $(A,B)$.
The corresponding generic notation for $\psi$ or $W$ is $\bzeta$.   
The columns of ${\bf A}$ and ${\bf B}$ are  denoted as $ \{ \bfA_1, \bfA_2, \bfA_3,\bfA_4\}$
and $ \{ \bfB_1, \bfB_2, \bfB_3,\bfB_4\}$, respectively. We stress that the bold characters are therefore associated with the generic
problem and not with a higher tensor order.
\\
%
\subsection{Coefficients of (ACE) and (DCE) as $4 \times 4$ determinants}
%

It has already been observed that the direct and adjoint problems share the same determinant, computed  using linear
combinations of columns, as
\begin{linenomath}
$$|\bfK| = \vert -dx {\bf B} + dy {\bf A} \vert =
  (-v ~dx  + u ~ dy)^2 (-v ~dx  + u ~ dy  + c ~dl)(-v ~dx  + u ~ dy  - c ~dl),   $$
\end{linenomath}
with
\begin{linenomath}
$$  c= \sqrt{\frac{\gamma p}{\rho}},\quad dl = \sqrt{dx^2 + dy^2},$$
\end{linenomath}
using the known eigenvalues of  the Euler flux Jacobian in an arbitrary direction \cite{JPDes_22}.
If $|\bfK|\neq0$, the Cauchy problem is well-posed, and the solution is expressed using the minors of $\bfK$.
 These minors are denoted as
 $\bfK^i_{jx} $ and $\bfK^i_{jy} $ with indices referring to the considered variable and the direction of
 differentiation. For example  $(\partial \bzeta_1/\partial x)$ reads
\begin{linenomath}
\beq
(\frac{\partial \bzeta_1}{\partial x})=
\frac{
\begin{vmatrix} 
  ~d\bzeta_1  & 0  & 0  &  0   & dy   &   0  &  0 & 0~ \\
  ~d\bzeta_2   & dx & 0  &  0  &  0   & dy   &  0 & 0 ~\\
  ~d\bzeta_3   & 0  & dx & 0  &  0   &   0  & dy & 0 ~\\
  ~d\bzeta_4   & 0  & 0  &  dx  &  0   &   0  & 0  & dy ~\\
  ~|   & |   &  |  &   |      &  |    &  |    &  |  & |~\\
  ~0 & \bfA_2 &  \bfA_3  & \bfA_4      &  \bfB_1  & \bfB_2 & \bfB_3 & \bfB_4 ~\\ 
  ~|   & |   &  |  &   |      &  |    &  |    & |   & |~\\
\end{vmatrix}
}{|\bfK|}
\label{e_solcaua}
\eeq
\end{linenomath}
\begin{linenomath}
\beq
(\frac{\partial \bzeta_1}{\partial x}) =
\frac{\bfK^1_{1x} d\bzeta_1 - \bfK^2_{1x} d\bzeta_2 + \bfK^3_{1x} d\bzeta_3 - \bfK^4_{1x} d\bzeta_4}{|\bfK|}
\label{e_solcaub}
\eeq
\end{linenomath}
Conversely, the generic Cauchy problem is ill-posed iff $|\bfK|=0$, that is  along the \st ,~ \cp  and \cm curves:
\bea
-v ~dx  + u ~ dy  &=& 0 \qquad \qquad  \textrm{\colb{\st~ streamtraces \qquad (all Mach numbers)}}   \label{e_ctraj}\\
-v ~dx  + u ~ dy  + c ~dl &=& 0 \qquad \qquad \textrm{\cm characteristics \quad\colb{(supersonic flow only)}}\label{e_ccp}\\
-v ~dx  + u ~ dy  - c ~dl &=& 0 \qquad \qquad \textrm{\cp characteristics \quad\colb{(supersonic flow only)}}\label{e_ccm}
\eea
In such case, the boundedness of $(\partial \bzeta_j/\partial x)$ $(\partial \bzeta_j/\partial y)$ along the characteristic curves and their
     expression outside these curves yields the (CE). Typically equation (\ref{e_solcaub}), that is valid outside the characteristic curves, yields the following (CE)
\begin{linenomath}
     $$\bfK^1_{1x} d\bzeta_1 - \bfK^2_{1x} d\bzeta_2 + \bfK^3_{1x} d\bzeta_3 - \bfK^4_{1x} d\bzeta_4 = 0$$
\end{linenomath}
 along the characteristic curves. Of course, seven other (CE) are derived from the boundedness of
 $(\partial \bzeta_2/\partial x)$... $(\partial \bzeta_4/\partial y)$ and we need to determine a minimal set of independant (CE)
  for each type of characteristic curve. 
 \\
 To that end, the $\bfK^i_{jx} $ and $\bfK^i_{jy} $ minors  
are expressed  as determinants of $4\times4$ matrices.
 From equations (\ref{e_solcaua}) and (\ref{e_solcaub}),
\begin{linenomath}
 $$ \bfK^1_{1x}
 =
 \begin{vmatrix} 
   dx & 0  &  0  &  0   & dy   &  0 & 0 ~\\
   0  & dx & 0  &  0   &   0  & dy & 0 ~\\
   0  & 0  &  dx  &  0   &   0  & 0  & dy ~\\
  ~ |   &  |  &   |      &  |    &  |    &  |  & |~\\
  ~ \bfA_2 &  \bfA_3  & \bfA_4      &  \bfB_1  & \bfB_2 & \bfB_3 & \bfB_4 ~\\ 
  ~ |   &  |  &   |      &  |    &  |    & |   & |~\\
\end{vmatrix}
$$ 
\end{linenomath}
 is the basic formula for  $\bfK^1_{1x}$. In all 32 corresponding expressions, the $dy$ terms of the first three lines 
 are eliminated by linear combinations of columns and the determinants are then expanded along these lines.
 \footnote{We assume that $dx \neq 0$ introducing $t=dy/dx$ and will further consider this hypothesis in the following. 
 For the sake of brevety we shall not discuss the simple specific case where $dx=0$.}
The minors arising in the (CE) derived from the existence of $(\partial \bzeta_1/\partial x)$ 
then read \\
\begin{linenomath}
\begin{minipage}{15cm}
\begin{align*} 
\bfK^1_{1x} &=  dx^3    ~&\vert \bfB_1    &~(\bfB_2-t \bfA_2)    &~  (\bfB_3-t \bfA_3) &~  (\bfB_4-t \bfA_4) &\vert  \\
\bfK^2_{1x} &= -dx^2 dy ~&\vert \bfA_2    &~~ \bfB_2 ~~          &~ (\bfB_3-t \bfA_3)  &~ (\bfB_4-t \bfA_4)  &\vert  \\
\bfK^3_{1x} &= -dx^2 dy ~&\vert \bfA_3    &~(\bfB_2-t \bfA_2)    &~~ \bfB_3   ~~       &~ (\bfB_4-t \bfA_4)  &\vert  \\
\bfK^4_{1x} &= -dx^2 dy ~&\vert \bfA_4    &~(\bfB_2-t \bfA_2)    &~ (\bfB_3-t \bfA_3)  &~~   \bfB_4    ~~    &\vert  \\
\end{align*}  
\end{minipage}
\end{linenomath}
\\and the corresponding expressions for $(\partial \bzeta_1/\partial y)$ are \\
\begin{linenomath}
\begin{minipage}{15cm}
\begin{align*}  
\bfK^1_{1y} &=& -dx^3  ~&\vert \bfA_1    &~(\bfB_2-t \bfA_2)    &~ (\bfB_3-t \bfA_3)  &~ (\bfB_4-t \bfA_4) &\vert  \\
\bfK^2_{1y} &=&  dx^3  ~&\vert \bfA_2    &~~ \bfB_2    ~~       &~ (\bfB_3-t \bfA_3)  &~ (\bfB_4-t \bfA_4) &\vert  \\
\bfK^3_{1y} &=&  dx^3  ~&\vert \bfA_3    &~(\bfB_2-t \bfA_2)    &~~ \bfB_3  ~~        &~ (\bfB_4-t \bfA_4) &\vert  \\
\bfK^4_{1y} &=&  dx^3  ~&\vert \bfA_4    &~(\bfB_2-t \bfA_2)    &~ (\bfB_3-t \bfA_3)  & ~~  \bfB_4    ~~   &\vert  \\
\end{align*}  
\end{minipage}
\end{linenomath}
\\Obviously
\begin{linenomath}
\beq
\bfK^2_{1x} = -t \bfK^2_{1y} \qquad \bfK^3_{1x} = -t \bfK^3_{1y} \qquad \bfK^4_{1x} = -t \bfK^4_{1y}.
\label{e_prop1}
\eeq
\end{linenomath}
Besides, expanding $| \bfK |$  along the first line of the matrix results in 
$$ |\bfK| = dx \bfK^1_{1x} +  dy \bfK^1_{1y}, $$
so that along a characteristic curve  $ \bfK^1_{1x} = - t \bfK^1_{1y} $. Corresponding equations are found for the other variables $\bzeta_j$ $j\geq 2$:
 $ \bfK^i_{jx} = - t \bfK^i_{jy}$ if $i\neq j$ whatever the value of $|\bfK|$ and   $ \bfK^j_{jx} = - t \bfK^j_{jy}$ if  $|\bfK|=0.$
If the minors are not all null, this proves that the four (CE) derived from the boundedness
of $(\partial \bzeta_k/\partial x)$ and their counterparts
derived from the boundedness of $(\partial \bzeta_k/\partial y)$ are proportional
so that only one set has  to be studied.
At this point, the case of the \st~ curves and the one of the \cp and \cm curves shall be discussed separately.
%
  \subsection{Number of independant (CE) along the \cp and \cm} 
%
 The abstract linear algebra point of view allows to calculate the number of independant (CE) in the specific case of the \cp and \cm curves.
 For $dx \neq 0$, $\bfK$ is easily found to be equivalent to
\begin{linenomath}
\beq
\begin{bmatrix}
   &    &    &    &    &    &   \\
   & dx \bfI &  &    & 0  &   &   \\
   &    &    &    &    &    &   \\
   &    &    &    &    &    &   \\
   &    &    &    &    &    &   \\
   & \bfA &   &    & \bfB -t \bfA  &  & \\
   &    &    &    &    &    &   
\end{bmatrix}
~~~\textrm{and ~then~ to~} \begin{bmatrix}
   &    &    &    &    &    &   \\
   & dx \bfI &  &    & 0  &   &   \\
   &    &    &    &    &    &   \\
   &    &    &    &    &    &   \\
   &    &    &    &    &    &   \\
   &    &  0 &    & \bfB - t \bfA  &  & \\
   &    &    &    &    &    &   
\end{bmatrix}
.
\eeq
\end{linenomath}
When the value $t$ is the one corresponding to a \cp or a \cm, for both the adjoint and the direct problem,
the rank of $ \bfB - t \bfA  $ is known to be 3 from the eigenanalysis of Euler equations flux Jacobians.
 Using classical theorems for the rank of block diagonal matrices
 and the rank of equivalent matrices, the rank of $\bfK$ is easily proven to be 7, one less that its size.
 In this case, the adjugate matrix is known to have rank one. This means that all (CE) (whose coefficients appear
 in the rows of the adjugate matrix) are proportional and reduced to only one independant equation.
 This result is very valuable as the corresponding explicit calculations from the agebraic expressions of the minors
 are very tedious. It has been used in \S 3.\\  
%
\subsection{Nullity of all minors along the \st~ curves}
%
The $4 \times4$ determinant involved in the expression of $\bfK^1_{1x}$ is
\beq
\vert \bfB_1    \quad (\bfB_2-t \bfA_2)   \quad  (\bfB_3-t \bfA_3) \quad  (\bfB_4-t \bfA_4) \vert.
\label{e_aofst}
\eeq
Along a streamtrace $t=dy/dx=v/u$ and $(\bfB-t \bfA)$ is a matrix of rank two.
As all its minors of rank three are null, the determinant (\ref{e_aofst}), calculated
by expansion along the first column, appears to be zero. To get the same property for
$\bfK^2_{1x}$,$\bfK^3_{1x}$ and $\bfK^4_{1x}$
a straighforward rewritting is needed.
For exemple, the determinant in $\bfK^2_{1x}$ reads
\begin{linenomath}
$$ \vert \bfA_2   \quad \bfB_2 \quad (\bfB_3-t \bfA_3) \quad (\bfB_4-t \bfA_4)  \vert ~~\textrm{or~ possibly}~~
\vert  \bfA_2   \quad (\bfB_2-t \bfA_2)  \quad (\bfB_3-t \bfA_3)  \quad (\bfB_4-t \bfA_4)  \vert.$$
\end{linenomath}
Under this last form, it is clearly null when the rank of $(\bfB-t \bfA)$ is two. Using the exact same arguments,
it is possible to prove
  that all  $\bfK^i_{jx}$ and  $\bfK^i_{ix}$ are null when  $t=v/u$.
  \\ 
   Besides, we may change our point of view and consider (\ref{e_aofst})  as a function of $t$ for fixed $W$.
   This function is obviously a polynomial of maximum degree three and $v/u$ is one of its roots.
   We hence expect $(t-v/u)$ to be a factor of all $\bfK^i_{jx}$, $\bfK^i_{jy}$ algebraic expressions. The explicit expressions of the coefficients
   for the (ACE) (see \cite{JPDes_22})
  and (DCE) (see Appendix A) confirm this property.
\\
Along the \st ~ curves, $\bfK$ is found to have rank 6 thanks to the technique used in the previous subsection.
As already stated, all
 $\bfK^i_{jx}$ and $\bfK^i_{jy}$ are zero and the adjugate matrix of $\bfK$ is the null matrix.
The derivation of the explicit (CE) along the trajectories then involves the multiplicity two of $(-v dx +u dy)$
and simplified  $\bar{\bfK}^i_{jx}$, $\bar{\bfK}^i_{jy}$ coefficients derived by removing $(-v dx +u dy)$ in the expressions of the corresponding
$\bfK^i_{jx}$, $\bfK^i_{jy}$. The proportionnality of the explicit (CE) derived from the boundedness of  $(\partial \bzeta_k/\partial x)$
and   $(\partial \bzeta_k/\partial y)$ is verified but is more easily presented based on the actual fomulas of the (CE)
  -- see \S 3 for the (DCE) and in \cite{JPDes_22} \S 2 for the (ACE). 
%
%
%
%
%
 \section{Derivation of characteristic equations for 2D inviscid compressible flows using conservative variables. }
%

\subsection{Flow properties, sets of variables, resulting equations for the usual derivation of the
  characteristic equations}
The classical derivations of the (DCE) use a set of Taylor expansions and mechanical equations for the variations
of a set of primitive variables.
These resulting equations are linear in the unknown derivatives of the primitive variables
 with non-linear functions of the state
 variables as coefficients. The variations are generally expressed in an orthonormal frame of reference
 $(\overrightarrow{\xi},\overrightarrow{\eta})$.
 This frame is defined by its angle with respect to the one induced by the local flow motion, that we
denote as ($\overrightarrow{t}$,$\overrightarrow{n}$)
 with $\phi$ the angle of $\overrightarrow{t}$ w.r.t. the $x$ axis. 
This approach leads to simpler expressions for the mechanical equations.
Some orientations of the vector $\vec{\xi}$ make the set of equations ill-posed:
$\overrightarrow{\xi}= \overrightarrow{t}$ for all flow regimes and
$ angle(\overrightarrow{\xi},\overrightarrow{t})=  \epsilon ~\textrm{arcsin}(1/M) ~~~\epsilon=\pm 1, $
if the flow is locally supersonic.
The curves where at each point the tangent vector has one of these specific orientations with respect to
the velocity are respectively the aforementioned streamtraces \st~, \cp (for $\epsilon=1$) and
\cm (for $\epsilon=-1$). Along these curves equations with only $\overrightarrow{\xi}$ derivatives
are satisfied. These ODEs are the classical (DCE).
 \\
  For a 2D flow of an ideal gas (\S 2.1), without any assumptions of constant stagnation enthalpy, 
  or constant entropy or null vorticity:
  (a) the stagnation enthalpy and entropy are found to be constant along streamtraces;
  (b) one (DCE) is found along the \cp \cm \cite{BonLun_89}.
  This (DCE), expressed in terms of variations of $\phi$, $\normU$,$S$ and $H$, reads   
\beq
\epsilon d \phi - \sqrt{M^2-1} \frac{d\normU}{\normU} - \sqrt{M^2-1} \frac{TdS - dH}{\normU^2}  = 0, 
\label{ebase}
\eeq  
where the Crocco equation can be used to rewritte the last term.
If $H$ is constant all over the fluid domain,
 the (DCE) also reads
\beq
\epsilon d \phi - \frac{\sqrt{M^2-1}}{1+ \frac{\gamma-1}{2}M^2} \frac{dM}{M} - \frac{\sqrt{M^2-1}}{\gamma r M^2}dS  = 0. 
\eeq
If $S$ is also constant, the flow is irrotational according to Crocco's theorem and the (DCE) reads
\beq
\epsilon d \phi - \frac{\sqrt{M^2-1}}{1+ \frac{\gamma-1}{2}M^2} \frac{dM}{M}  = 0. \label{esim} 
\eeq 
This equation can be integrated involving the Prandtl-Meyer function,
\begin{linenomath}
   $$ \nu(M) =\sqrt{\frac{\gamma+1}{\gamma-1}}
      \tan^{-1} \left( \sqrt{\frac{\gamma-1}{\gamma+1} (M^2 -1)} \right) -\tan^{-1}(\sqrt{M^2-1}),$$ 
\end{linenomath}
and the final algebraic equations resulting from the integration of (\ref{esim}) are      
\beq
 k^- = \phi + \nu(M)~~ \textrm {is constant along a}~~ {\cal C}^- ~~~~~~~~~
      k^+ = \phi - \nu(M)~~ \textrm {is constant along a}~~ {\cal C}^+. \label{eint}
\eeq
Other sets of primitive variables may be used to derive (\ref{ebase}) and the resulting simplified
equations \cite{MeyGol_48,Del_08}. Also possible is the demonstration of these equations in a mapping of
 the plane rather than in the aforementioned frame of reference \cite{MeyGol_48}.
\\
In case the flow is assumed to be irrotational from the beginning, a very fast demonstration
is possible involving the velocity potential -- see \cite{And_03} or \cite{Sha_54,LieRos_56,Gro_03}.
The governing nonlinear equation for
a two-dimensional potential flow is:
\beq
(1-\frac{u^2}{c^2})\frac{\partial u}{\partial x} - \frac{2uv}{c^2}\frac{\partial u}{\partial y} + (1- \frac{v^2}{c^2})\frac{\partial v }{\partial y} =0
\eeq
and of course 
\begin{linenomath}
$$ du = \frac{\partial u}{\partial x}dx + \frac{\partial u}{\partial y}dy   ~~~~~~~
 dv = \frac{\partial v}{\partial x}dx + \frac{\partial v}{\partial y}dy .  $$
\end{linenomath}
 This results in a simple $(3\times3)$ linear system for $(\partial u/\partial x)$, $(\partial v/ \partial y)$
and $(\partial u/\partial y)$ (that is equal to $(\partial v/\partial x)$ for the potential flow) from which
(\ref{esim}) and (\ref{eint}) are easily derived.
 %
%
\subsection{Derivation of the characteristic equations in conservative variables }
 The Cauchy problem aims at computing the derivatives $(\partial W/ \partial x)$ $(\partial W/ \partial y)$ from the values
 of $W$ at two neighboring points $a$ and $b$.  From the Euler
 equations and basic first order Taylor formulas, the equations of this problem read 
\begin{linenomath}
\begin{equation} 
\begin{bmatrix}
dx & 0  & 0  & 0  & dy & 0  & 0  & 0 \\
0  & dx & 0  & 0  & 0  & dy & 0  & 0 \\
0  & 0  & dx & 0  & 0  & 0  & dy & 0 \\
0  & 0  & 0  & dx & 0  & 0  & 0  & dy \\
   &    &    &    &    &    &    &   \\
   &    &    &    &    &    &    &   \\
   &    & A  &    &    &    & B  &   \\
   &    &    &    &    &    &    &   
\end{bmatrix}
\begin{bmatrix}
(\partial \rho /\partial x) \\
(\partial \rho u/\partial x) \\
(\partial \rho v/\partial x) \\
(\partial \rho E/\partial x) \\
(\partial \rho /\partial y) \\
(\partial \rho u/\partial y) \\
(\partial \rho v/\partial y) \\
(\partial \rho E/\partial y) 
\end{bmatrix}
=
\begin{bmatrix}
 d\rho \\
 d\rho u \\
 d\rho v \\
 d\rho E \\
 0 \\
 0 \\
 0\\
 0
\end{bmatrix}
\label{e_cbase}
\end{equation}
\end{linenomath}
 with $(dx,dy)=\overrightarrow{ab}$, $( d\rho ,  d\rho u ,  d\rho v ,  d\rho E ) =
(\rho^b - \rho^a,  \rho u^b - \rho u^a,  \rho v^b - \rho v^a, \rho E^b - \rho E^a)  $
and $A$,$B$ the Euler flux Jacobian matrices.
The determinant is easily calculated as 
\begin{linenomath}
\begin{equation}
  D = 
\begin{vmatrix}
      &    &    &    &    &    &   \\
      & dx I&    &    &  & dy I  &   \\
      &    &    &    &    &    &   \\
      &    &    &    &    &    &   \\
      &    &    &    &    &    &   \\
      & A  &    &    &  & B  &   \\
      &    &    &    &    &    &   
\end{vmatrix}
= 
\begin{vmatrix}
   &    &    &    &    &    &   \\
   & dx I &  &    & 0  &   &   \\
   &    &    &    &    &    &   \\
   &    &    &    &    &    &   \\
   &    &    &    &    &    &   \\
   & A&   &    & B-(dy/dx) A  &  & \\
   &    &    &    &    &    &   
\end{vmatrix}
= dx^4\vert B - (dy/dx) A \vert 
\end{equation}
\end{linenomath}
Of course, $D = \vert dx B - dy A \vert = \vert  -dx B + dy A \vert $
 and the problem is ill-posed along the \st ,~ \cp ,~\cm curves as recalled in \S 2.3.
The fact that the Cauchy problem is ill-posed along these specific curves whereas $(\partial W/\partial x)$ and
$(\partial W/\partial y)$ are actually defined and bounded,
implies that not only the denominator
in the Cramer formula applied to (\ref{e_cbase}) is equal to zero but also the eight numerators corresponding
to the unknowns  $(\partial \rho /\partial x)$, $(\partial \rho u/\partial x)$, $(\partial \rho v /\partial x)$,
$(\partial \rho E /\partial x)$ $(\partial \rho /\partial y)$,  $(\partial \rho u /\partial y)$,
$(\partial \rho v/\partial y)$, $(\partial \rho E/\partial y)$ must be equal to zero.
This is precisely the principle of the method of caracteristics that allows to derive ODEs along
the caracteristic curves.
\newline
 The Euler flux Jacobian matrices in $x$ and $y$ direction, $A$ and $B$, are equal to
\begin{linenomath}
$$
A = 
\begin{bmatrix} 
	0  &  1  & 0 & 0\\
	\gamma_1 E_c -u^2  & (3-\gamma)u &  - \gamma_1 v  & \gamma_1 \\
	-uv   & v &  u  &0 \\
	(\gamma_1 E_c - H)u  & H-\gamma_1 u^2 &  - \gamma_1 uv  & \gamma u\\
\end{bmatrix}
~~~~~~~~
B = 
\begin{bmatrix} 
	0  & 0 & 1 & 0\\
	-uv  &  v  &  u & 0 \\
	\gamma_1 E_c -v^2  &  -\gamma_1 u  & (3-\gamma)v  & \gamma_1\\
	(\gamma_1 E_c - H)v  &  -\gamma_1 uv  & H-\gamma_1 v^2       & \gamma v \\
\end{bmatrix}
$$
\end{linenomath}
(with $\gamma_1 = \gamma -1$) and the following notations are introduced: 
$$ t = \frac{dy}{dx} ,\quad \kappa = u t -v.$$
The columns of Jacobian are denoted by $A_1$ to $A_4$
and $B_1$ to  $B_4$ so that
$ A = [A_1|A_2|A_3|A_4] $,
$ B = [B_1|B_2|B_3|B_4] $.
The the principle of the calculation of the ODE satisfied along the \st, \cp and \cm curves
is recalled for variable $(\partial \rho /\partial x)$ which boundedness requires that
\begin{linenomath}
\begin{equation} 
\begin{vmatrix} 
  ~d\rho  & 0  & 0  &  0   & dy   &   0  &  0 & 0~ \\
  ~d\rho u   & dx & 0  &  0  &  0   & dy   &  0 & 0 ~\\
  ~d\rho v   & 0  & dx & 0  &  0   &   0  & dy & 0 ~\\
  ~d\rho E   & 0  & 0  &  dx  &  0   &   0  & 0  & dy ~\\
  ~|   & |   &  |  &   |      &  |    &  |    &  |  & |~\\
  ~0 & A_2 &  A_3  & A_4      &  B_1  & B_2 & B_3 & B_4 ~\\ 
  ~|   & |   &  |  &   |      &  |    &  |    & |   & |~\\
\end{vmatrix}
 = 0.
\end{equation}
\end{linenomath}
The determinant is expanded along the first column with the notations of next equation:
\begin{linenomath}
\begin{equation}
 K_{1x}^1 d\rho - K_{1x}^2 d\rho u + K_{1x}^3 d\rho v - K_{1x}^4 d\rho E = 0.
\label{e_cdrodx}
\end{equation}
\end{linenomath}
In this equation, for example 
\begin{linenomath}
\beq
  K_{1x}^4=
\begin{vmatrix} 
  ~0   & 0  & 0    &  dy   &   0   &  0  & 0~ \\
  ~dx   & 0   & 0   &  0   &   dy    & 0  & 0 ~\\
  ~0   & dx   & 0    &  0   &   0    &  dy  & 0 ~\\
  ~|   & |   &  |   &  |    &  |    &  |  & |~\\
  ~A_2 & A_3 &  A_4 &  B_1  & B_2 & B_3 & B_4 ~\\ 
  ~|   & |   &  |   &  |    &  |    & |   & |~\\
\end{vmatrix}
=
\begin{vmatrix} 
  ~0   &  0  & 0   &  dy   &   0  &  0 & 0~ \\
  ~dx   & 0    & 0   &  0   &   0  &  0 & 0 ~\\
  ~0   & dx    & 0  &  0   &   0  &  0 & 0 ~\\
  ~|   & |   &  |   &  |    &  |     &  |  & |~\\
  ~A_2 & A_3 &  A_4 &  B_1   & B_2-t A_2 & B_3 -t A_3 & B_4 ~\\ 
  ~|   & |   &  |   &  |    &  |    & |   & |~\\
\end{vmatrix} 
\eeq
\end{linenomath}
Finally
\begin{linenomath}
\begin{equation}
 K_{1x}^4 = dx^2 dy ~ \vert~~ A_4~~ (B_2-t A_2) ~~ (B_3- t A_3)~~ (B_4) ~~\vert =
 dx^2 dy ~ \kappa ~ \gamma_1  ~(tv+u)
\end{equation}
\end{linenomath}
At this step, no assumption on the value of $t$ with respect to the velocity vector $(u,v)$.
The other terms of the differential form are equal to 
\begin{linenomath}
\begin{eqnarray}
 K_{1x}^1 &=& - dx^3 ~ \kappa ~ (v ~ \kappa^2 -  \gamma_1 ~ v ~ (1+t^2)~ H + \gamma_1 ~ (u t +v)~ E_c + 2 ~ \gamma_1 ~ t^2 ~ v ~ E_c) \\
 K_{1x}^2 &=& -dx^2 ~ dy ~ \kappa ~ (\gamma ~ \kappa ~ v + 2 ~ \gamma_1 E_c) \\
 K_{1x}^3 &=&   dx^2 ~ dy ~ \kappa ~ (-2 ~ t ~E_c + \gamma ~ v ~ (t ~ v+u))
\end{eqnarray}
\end{linenomath}
 The characteristic equation (\ref{e_cdrodx}) can hence be rewritten as  
\begin{linenomath}
\begin{eqnarray}
&-& dx^3 ~ \kappa ~ (v ~ \kappa^2 -  \gamma_1 ~ v ~ (1+t^2) H + \gamma_1 ~ (u t +v)~ E_c + 2 ~ \gamma_1 ~ t^2 ~v~  E_c )~ d\rho \nonumber \\
&+&dx^2 ~ dy ~ \kappa ~ (\gamma ~ \kappa ~ v + 2 ~ \gamma_1~ E_c)~ d\rho u +dx^2 ~ dy ~ \kappa ~ (-2 ~ t~ E_c + \gamma ~ v ~ (t ~ v+u))~ d\rho v\nonumber \\ 
&-& dx^2 dy ~ \kappa ~ \gamma_1  ~(tv+u)~ d\rho E = 0
\end{eqnarray}
\end{linenomath}
 If $dx \neq 0$ and $\ww \neq 0$, this equation may be further simplified: 
\begin{linenomath}
\begin{eqnarray}
&-&  (v ~ \kappa^2 - \gamma_1 ~ v ~ (1+t^2) H + \gamma_1 ~ (u t +v) E_c + 2 ~ \gamma_1 ~ t^2 ~ v~ E_c )~ d\rho \nonumber \\
&+& t ~ (\gamma ~ \kappa ~ v + 2 ~ \gamma_1~E_c)~ d\rho u +t ~  (-2 t ~ E_c  + \gamma ~v ~ (t ~ v+u))~ d\rho v\nonumber \\ 
  &-& t ~ \gamma_1  ~(tv+u)~ d\rho E = 0
  \label{e_drhodx2}
\end{eqnarray}
\end{linenomath}
Of course, along the \st ~curves, $\kappa$ is null and the equation (\ref{e_drhodx2}) seems to be only valid for the \cp and
\cm curves. Actually as $(-vdx+udy = \ww dx)$ has a multiplicity of two
in the determinant $D$, (\ref{e_drhodx2}) is also needed for the existence
of $(\partial \rho / \partial x)$ along
 the \st~ curves (see \S 3.3).\\
 For the other seven partial derivatives, the expression of the coefficents
 of the (CE) are presented in the Appendix A. Besides, it is proven in \S 2 that, $W_l$ being
  one of the four conservative variables, the equations for the boundedness of
 $(\partial W_l/\partial x) $ and $(\partial W_l/\partial y) $
 along the characteristics are proportionnal by a $-t$ factor. 
 Therefore, only the counterpart of equation (\ref{e_drhodx2}) for the existence of
 $(\partial \rho u / \partial x)$, $(\partial \rho v / \partial x)$ and $ ~(\partial \rho E / \partial x)$
are presented hereafter:
\begin{linenomath}
\bea
  &-& ~\gamo ~ t ~ E_c  (-\gamo ~H   + (\gamma +1)~E_c) ~d \rho \nonu \\
  &+& (\gamo (\gam t u+ \ww) E_c  +v^2 ~ \ww + \gamo ~ (v -\gamo ~ t ~ u) H ) ~ d \rho u
\nonu \\
&-& t ~ ( u^2 \ww + \gamo (\ww - \gam v) E_c  +  \gamo (\gamo v- u t) H) ~d \rho v
\nonu \\ 
&+& t ~ \gamo (\gamo H  - (\gamma +1) E_c) ~d \rho E =0
\label{e_drhoudx2}
\eea  
\end{linenomath}
\begin{linenomath}
\bea
  &+& \gamo ~ E_c ~ t^2   (\gamo ~ H - (\gamma +1) ~ E_c) ~d\rho \nonu \\
   &-&  t(-v^2 \ww - \gamo \ww E_c - \gam \gamo t u E_c  + \gamo(\gamo t u-v)H ) ~d \rho u \nonu \\
&+& (- v^3 + 3 t u v^2 - 2 t^2 u^2 v + \gamo((2-\gam)t^2v - \ww) H +  \gamo (\gam t^2 v +\ww) E_c ) ~d \rho v  \nonu \\
  &-& t^2 \gamo (-\gamo ~ H + (\gamma +1) E_c)~d \rho E=0
\label{e_drhovdx2}
\eea
\end{linenomath}
\begin{linenomath}
\bea
  &-&   \gamo t(u+tv) E_c ( \gamo E_c+ (2-\gamma)H)~ d \rho \nonu \\
  &+& t(\gamo u^2 + v \ww + \gamo t u v)(\gamo E_c + (2-\gamma)H)~ d\rho u \nonu \\
  &-& t(u \ww - \gamo t v^2 - \gamo u v)(\gamo E_c + (2-\gamma)H)~ d \rho v \nonu \\
  &+& ( - v \ww^2 - \gamo ((1+\gam t^2)v + \gamo t u) E_c + \gamo^2 t(u+tv)H -\gamo \ww H)~ d \rho E = 0
\label{e_drhoEdx2}
\eea
\end{linenomath}
%
%
\subsection{Ordinary differential equations along the streamtraces \st}

Along the trajectories, $u dy - v dx =0$ is a zero of the denominator of Cramer's formulas with a multiplicity of two. 
The term $\ww dx = u dy - v dx$ appears to be
a factor of all $K^l_{mx}$ and the expressions obtained by removing this term
 from $K^l_{mx}$ is denoted by $\bar{K}^l_ {mx}$. 
Obviously wherever the Cauchy problem is well-defined 
\begin{linenomath}
\beq
\frac{\partial \rho}{ \partial x} =
\frac{\bar{K}^1_ {1x} d\rho - \bar{K}^2_ {1x} d\rho u+ \bar{K}^3_ {1x} d\rho v - \bar{K}^4_ {1x} d\rho E}
     { (-v ~dx  + u ~ dy) (-v ~dx  + u ~ dy  + c ~dl)(-v ~dx  + u ~ dy  - c ~dl)}.
\label{e_n1}
     \eeq
\end{linenomath}
 If the point $a$ is fixed and the neighboring point  $b$ is moved closer and closer to \st$_a$,
  $(-v ~dx  + u ~ dy) \rightarrow 0$ and the boundedness of $ (\partial \rho / \partial x)$  
 requires the numerator of (\ref{e_n1}) to be equal to zero:
\beq
\bar{K}^1_ {1x} d\rho - \bar{K}^2_ {1x} d\rho u+ \bar{K}^3_ {1x} d\rho v - \bar{K}^4_ {1x} d\rho E =0.
\eeq
The expressions for the $\bar{K}$ coefficients are much simpler than those of the $K$ coefficients thanks to the nullity of $\ww$. 
It can be verified by hand or by formal calculation that the (CE) expressing the boundedness
of $(\partial W_l/\partial x)$ and their counterparts for $(\partial W_l/\partial y)$ are propotional by a $-t$ factor.
It is also possible to check that  $\bar{K}^l_{lx} = - t \bar{K}^l_{ly} $ and then use
the continuity of the $\bar{K}$ coefficients as functions of  $(W,dx,dy,\gam)$
 and the validity of  $\bar{K}^l_{mx} = - t \bar{K}^l_{my} ~~l<>m $ outside the variety $u dy -v dx = 0$.
 We denote  $(\gamo E_c + (2-\gam)H)$  as $ G$; The fully simplified differential forms then read 
\begin{linenomath}
\begin{eqnarray}
 (2E_c -H)~ d\rho - u~ d\rho u  -v ~d\rho v + d\rho E &=& 0 \label{e_p2}\\
  ((\gamma +1) E_c-\gamo H)~ (E_c d\rho+d \rho E)  - (\gamma E_c + (2-\gamma)H)~(u d \rho u + v d\rho v)  &=&0 \label{e_p3}\\
 - E_c G  ~d\rho   + u G ~d\rho u  + v G  ~d\rho v  +  (\gamo H - \gamma E_c)~ d \rho E &=&0 \label{e_p4},
\end{eqnarray}
\end{linenomath}
 where equation (\ref{e_p3}) is obtained twice, from the existence of the derivatives of both momentum components. 
 The rank of the matrix was first calculated with Maple software and found unsurprisingly to be two. It can
also be verified by hand calculation that $Ec \times $  (\ref{e_p2}) minus (\ref{e_p3}) is exactly equation (\ref{e_p4})
 and then, finally, that the $ d\rho$ and $d \rho E$ columns in \{(\ref{e_p2}),(\ref{e_p3})\} are not linked.
\\
Along the streamtrace \st, the derivative of the total enthalpy, $H$, and the entropy, $S$, are expected to be null
 \cite{And_03}.
Before verifying that these properties are ensured by the (CE) found just before, the differential of   $p/\rho^\gam$
 times $\rho^\gam$ and the differential of $H$ times $\rho$ are calculated:
\begin{linenomath}
\begin{eqnarray}
  \rho^\gam d(p/\rho^\gam) &=& (2E_c -H)~ d\rho - u~ d\rho u  -v ~d\rho v + d\rho E   \label{e_q1} \\
  \rho d H &=& (-\gamma E + 2(\gamma -1) E_c) d\rho -\gamo u d\rho u - \gamo v d\rho v + \gamma d\rho E \label{e_q2} 
\end{eqnarray}
\end{linenomath}
 Equation (\ref{e_p2}) is exactly the same as  $ \rho^\gam d(p/\rho^\gam) = 0 $.
This yields  $d(p/\rho^\gam) = 0 $ and obviously
the derived (CE) contain the property of constant entropy along the streamtraces.
Besides $ G \times $ (\ref{e_p2}) plus (\ref{e_p4}) yields, after simplification by a common factor,  
 $ -G d\rho + d\rho E = 0  $.
Finally adding this equation to $\gamo$ times equation (\ref{e_p2}), precisely yields $\rho dH = 0 $,
so that second expected property for Euler flows along the trajectories is contained in the derived (CE).
%
%
%
\subsection{Ordinary differential equations along the \cp \cm}
The proportionality of all eight differential forms along the \cp, \cm has been proven in \S 2.4.
and equation (\ref{e_drhodx2}) may be retained as the relevant (DCE). The factor of $d \rho$
is quite complex in (\ref{e_drhodx2}) but it admits a simpler form derived from the degree two equations
 satisfied by $t^+$ and $t^-$: these curve slopes are defined as
\begin{linenomath}
\beq
\colt{t^\pm} = \tan(\phi + \beta)~~~~ \textrm{with}~~~ \tan \phi = \frac{v}{u} ~~~ \sin \beta = \pm \frac{1}{M},
\label{etapb}
\eeq
\end{linenomath}
or alternatively by their explicit expressions \cite{JPDes_22}:
\begin{linenomath}
\beq
t^\pm = \frac{uv \pm c\sqrt{u^2 + v^2 -c^2} }{u^2-c^2}.
\label{eteps} 
\eeq
\end{linenomath}
 They are the two roots of the following degree two equation \cite{JPDes_22}
\begin{linenomath}
\beq
 \gamma_1 ~(1+t^2)~ H =  \gamo~(1+ t^2)~E_c + (tu -v)^2,
 \label{e_deg2}
\eeq
\end{linenomath}
 which allows to simplify the expression of (\ref{e_drhodx2}) along a \cp or a \cm. The final (DCE) reads
\begin{linenomath}
\beq
\gamo  E_c (u+tv) d\rho -(\gamma u (u+tv)- 2 E_c) d\rho u -(\gamma v (u+t v)- 2 E_c t) d\rho v
+ \gamo(u+t v) d\rho E = 0
\label{e_dcpcm}
\eeq
\end{linenomath}
 with $t=t^+$ for a \cp and by $t=t^-$ for a \cm.
\\
It has been first checked  that this \cp (resp. \cm) differential form is linked
with $\{dk^+, dH, dS\}$ (resp. $\{dk^-, dH, dS\}$) consistently with equation (\ref{ebase}).
 It proven in Appendix B that these (CE), expressed with different variables, are the same.
%

\section{Assessment of the direct and adjoint characteristic equations for a nozzle flow}

\subsection{Supersonic nozzle configuration}
\begin{figure}[htbp]
  \begin{center}
	  \includegraphics[width=0.95\linewidth]{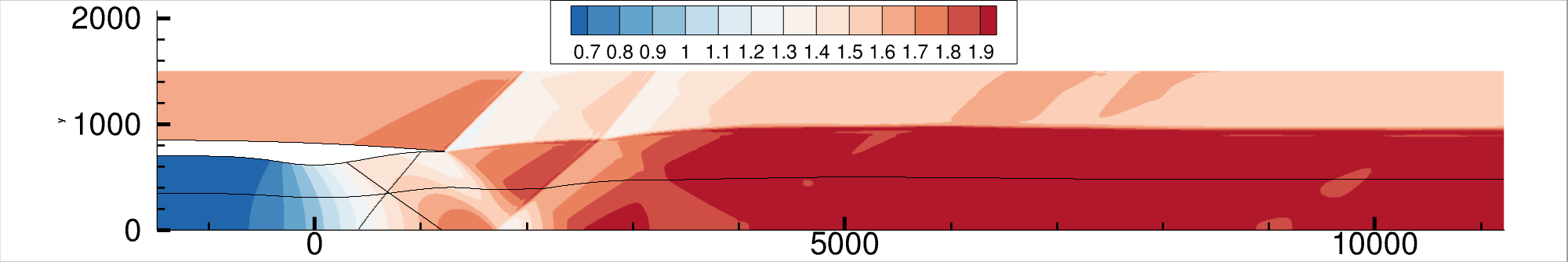}
	  \caption{ SENECA nozzle. Iso-Mach number lines. Selected \st, \cp and \cm for the validation of the (DCE) and (ACE)}
\label{fig:geoflow}
\end{center}
  \end{figure}

The test case for the assessment of the (DCE) and (ACE) is a supersonic nozzle designed for an  aircraft flying at Mach 1.6.
The geometry was designed in the framework of the SENECA EU project \cite{Mou_22}.
The nozzle is a classical convergent-divergent duct. At the subsonic injection plane, the stagnation conditions are fixed
and the ratio of the farfield flow static pressure (resp. temperature) to the total pressure (resp. total temperature) at the inlet
 are $p_{i_{inl}}/p_{i_\infty}=1.605$ and $T_{i_{inl}}/T_{i_\infty}=1.606$.
 As an inviscid flow is calculated, these two conditions and the farfield-Mach number fully define the flow.
 \\
 The original geometry was designed for a circular axisymmetric engine and the nozzle contour in the symmetry plane was simply extracted
 to define the 2D geometry. Consequently the mass flow rates are different. Due the different laws of areas along the $x$ coordinate (the inlet and
 main flow direction), the flows are also different along the midline of the nozzle. Nevertheless,
 the Mach number distribution along this line are similar with $M=1$ at the throat and $M\simeq 2.1$ at the exit.
 \\
 The fonctional output of interest is the thrust $T$,
 \begin{linenomath}
 \beq
 T = \int_{\Gamma_w} (p-p_\infty) ds_x + \int_{\Gamma_{inl}} (\rho u^2+p) ds_x, \label{e_thru}
 \eeq
 \end{linenomath}
 where $\Gamma_w$ is the internal wall of the nozzle, $\Gamma_{inl}$ the injection section, and $s_x$ the $x$-component of the surface vector
 (oriented towards the outside of the fluid domain).\\
 The nozzle and the fluid domain are presented in figure \ref{fig:geoflow} as well as the Mach-number distribution and the
 \st, \cp and \cm curves selected for the (CE) assessment.
 A structured 8-block mesh with 320000 cells discretizes the fluid domain.   
 The flow simulations are run with the elsA code \cite{CamHeiPlo_13}
 using the Jameson-Schmidt-Turkel (JST) scheme \cite{JamSchTur_81}. The adjoint calculation for the thrust is
 run with the  discrete adjoint module
 of the  $elsA$ code \cite{JPRenDum_15} which
  is used for both research \cite{JPDruPha_04,JPNguTro_12,TodVonBou_16,JPRenLab_22}
 and application \cite{CarDesDum_14,MehDesBen_15,MehDumCar_16} purposes. In \cite  {JPRenLab_22}, it is proven that
 the discrete adjoint of the (JST) scheme on a structured mesh is dual consistent for Euler flows
 but in the cells next to those adjacent to a boundary. This adjoint consistency property inside the fluid domain
  is very valuable to numerically assess continuous adjoint properties. 
 The change in the Jacobian to get the dual consistency
 in the vicinity of the physical boundaries is complex to implement in an industrial code. In this study,
 it is not used and slight oscillations of the adjoint field are observed near the boundaries.
 For the sake of simplicity, considering the mesh density,
 the points inside the first two cells in the vicinity of a boundary are removed from the extracted charateristic curves.
 The first component of the adjoint field is presented in figure \ref{fig:geoadj}. The exact adjoint of the thrust is expected
 to be zero downstream the backward \cp characteristic (and corresponding \cm in the symmetric non-plotted domain)
 emanating from the rear of the nozzle, as no perturbation downstream these lines can affect the  flow on the
 support of $T$ (We recall that the velocity is supersonic in the divergent part of the nozzle). This property is well satisfied by
  the discrete adjoint.
 \\
 A series of recent publications have studied the properties the Euler lift- and drag-adjoint
 fields, in particular in the vicinity of the wall
 \cite{Loz_17,Loz_18,Loz_19,JP_20,JPRenLab_22}. One of the remarkable properties of these
 fields when associated to a wall-pressure integral
 is the proportionality of $\psi_1$ and $\psi_4$ by a factor $H$ \cite{GilPie_01}
 which  is numerically very well satisfied \cite{JPRenLab_22}.
 In addition to the validation of the paper results, we take the opportunity provided by
 the specific function of interest (\ref{e_thru})
 (that does not only depend on the static pressure), to plot together
 $\psi_1$ and $H \psi_4$ -- see figure \ref{fig:geoadj}. The two quantities seem similar
 and futher studies may be devoted to the conditions on the output functional of interest
  for which $\psi_1 = H \psi_4 $ should be observed.
\begin{figure}[htbp]
  \begin{center}
	  \includegraphics[width=0.95\linewidth]{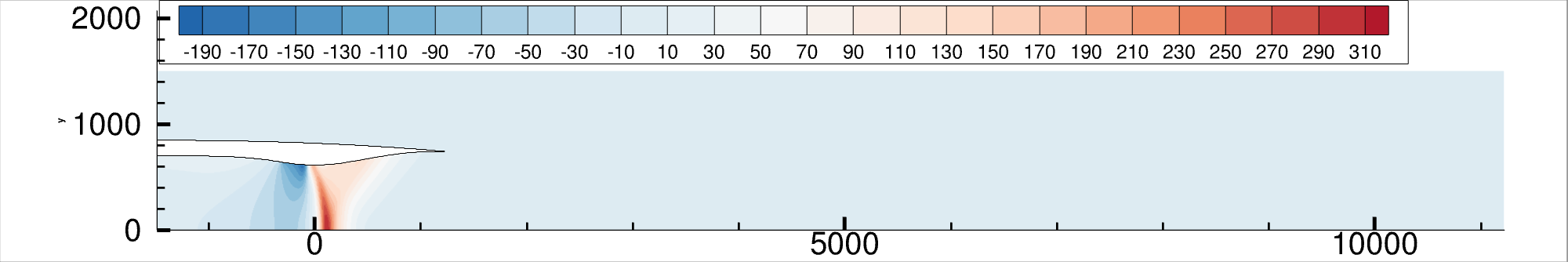}
	  \caption{ SENECA nozzle. Iso lines of the discrete adjoint of the thrust, component $\psi_1$ (dual of the mass-flow residual)}
\label{fig:geoadj}
\end{center}
\end{figure}
\begin{figure}[htbp]
  \begin{center}
	  \includegraphics[width=0.8\linewidth]{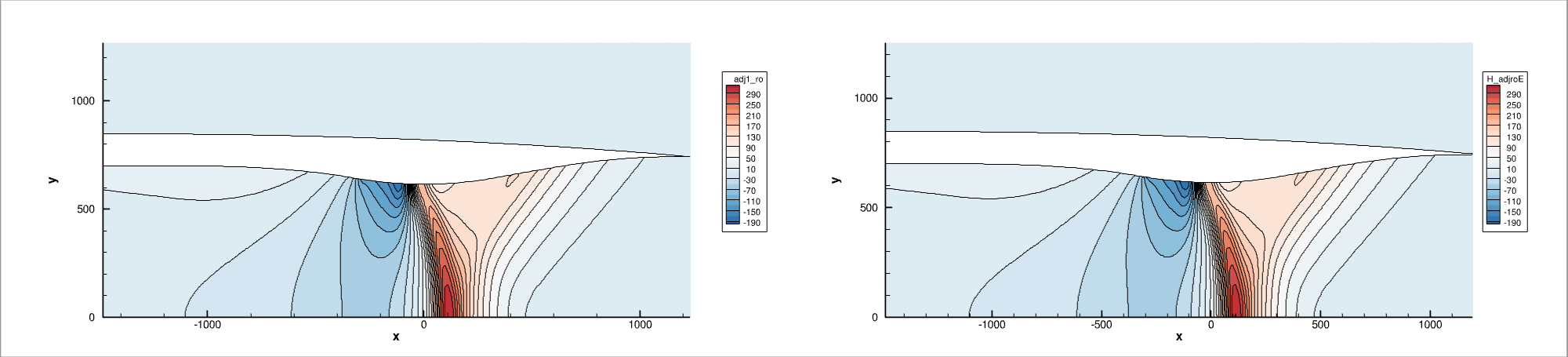}
	  \caption{ SENECA nozzle. Iso-$\psi_1$ (left) and iso-(H$\times\psi_4$) with same levels}
\label{fig:geoadj}
\end{center}
  \end{figure}
  
%
\subsection{Assessment of the (CDE)}
\begin{figure}[htbp]
  \begin{center}
	  \includegraphics[width=0.4\linewidth]{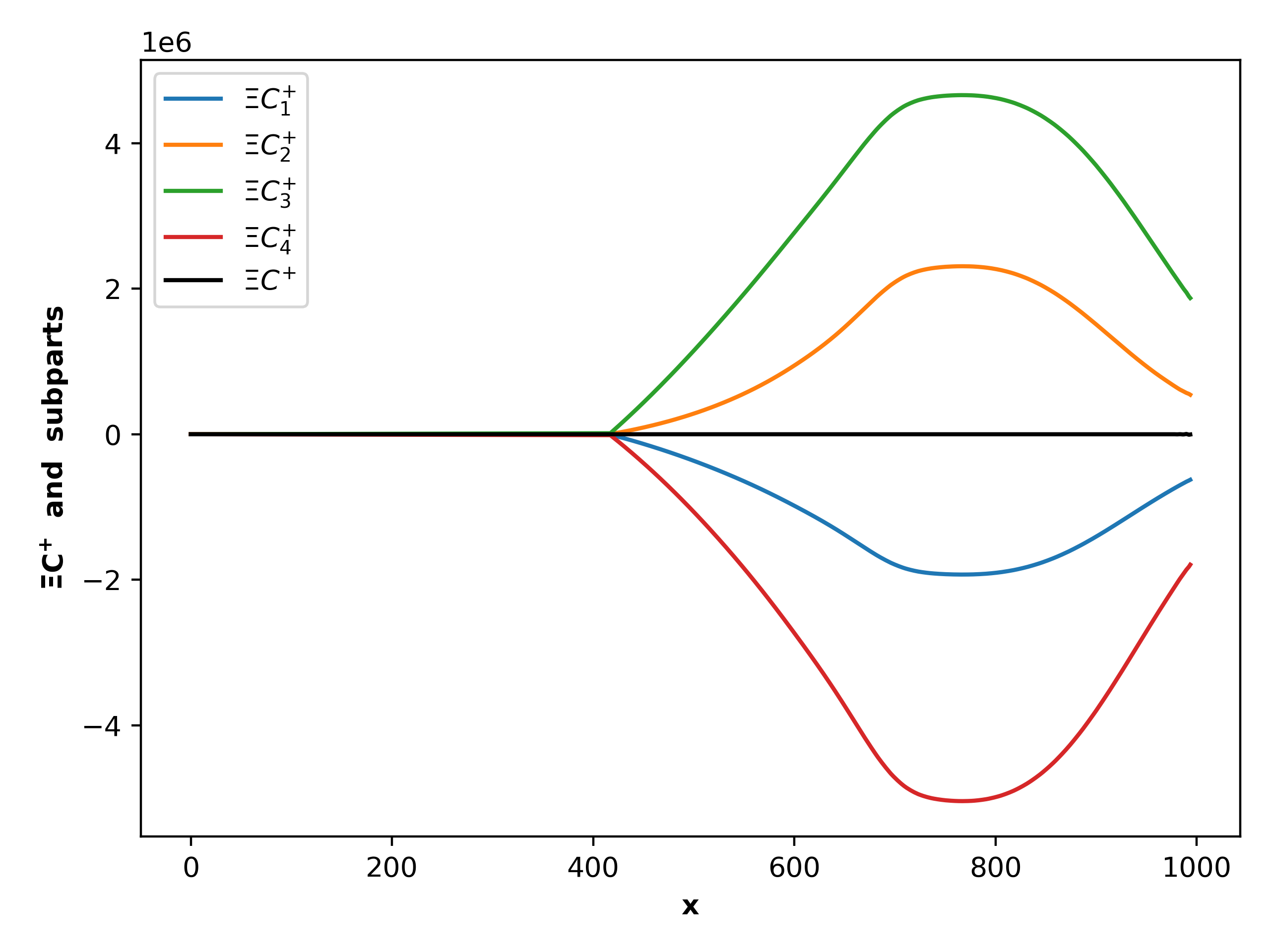}
	  \includegraphics[width=0.4\linewidth]{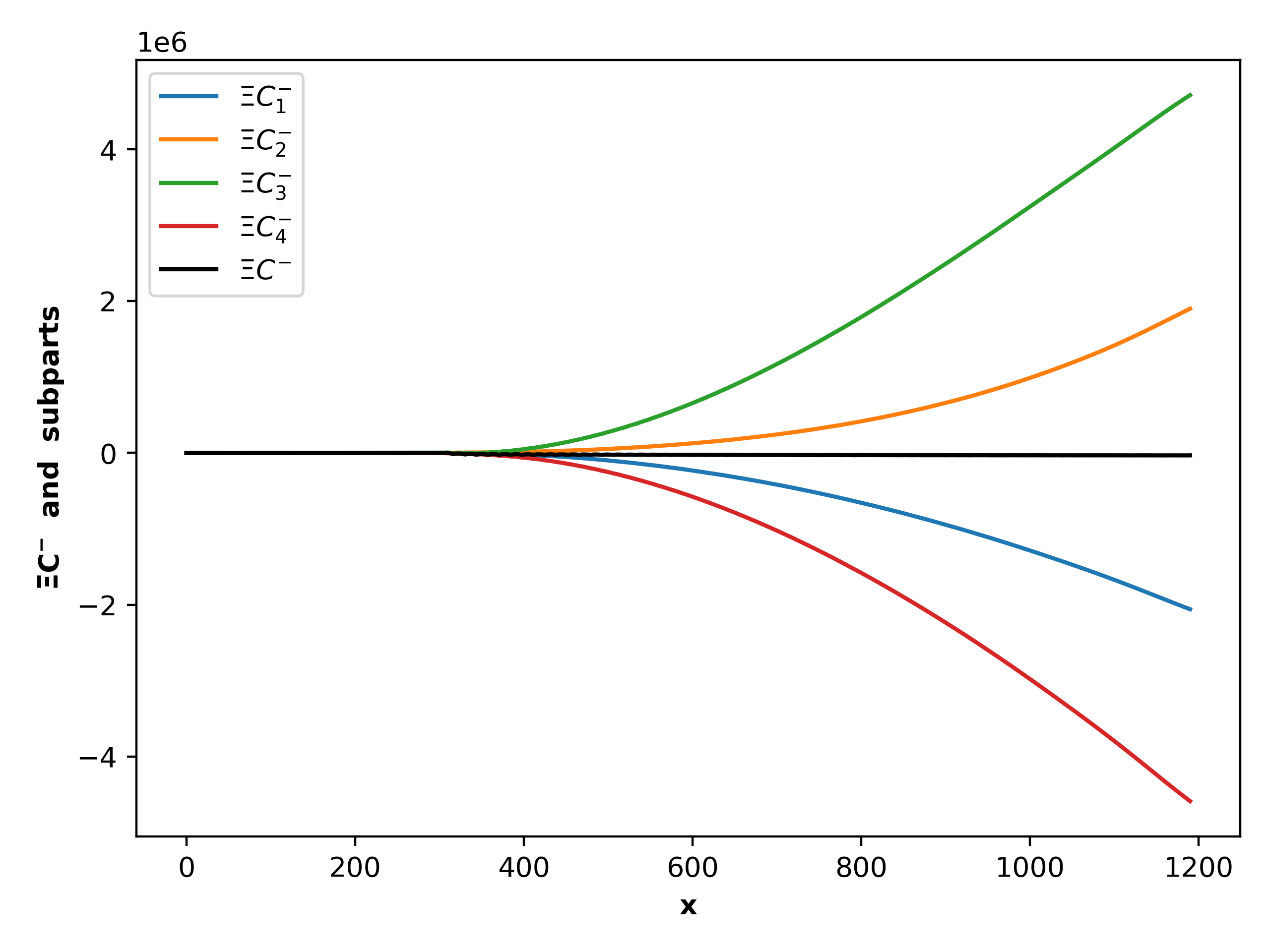}
	  \caption{ Numerical assessment of (DCE) (\ref{e_dcpcm}).
      $\Xi C^+$ (left), $\Xi C^-$ (right) and their subparts integrated along the selected \cp \cm.
    Method of verification: the black curve should ideally coincide with the $x$ axis. }
\label{fig:dCpmflow}
\end{center}
\end{figure}

The numerical assessment method consists in integrating the (CDE) along the characteristic curves for
the aforementioned fine mesh flow. More precisely, for the \cp and \cm, the following integrals are calculated
\begin{linenomath}
{\footnotesize
\bea
\Xi C^+ &=& \int_{{\cal C}^+}  \left( \gamo E_c (u+vt^+) ~\frac{d\rho}{ds} - (\gamma u(u+vt^+)-2 E_c ) \frac{d\rho u}{ds}  
- (\gamma v(u+vt^+)- 2E_c t^+) \frac{d\rho v}{ds} +\gamo (u+v t^+)\frac{d\rho E}{ds}  \right) ds \nonumber\\
\Xi C^- &=& \int_{{\cal C}^-}  \left( \gamo E_c (u+vt^-) ~\frac{d\rho}{ds} - (\gamma u(u+vt^-)-2 E_c ) \frac{d\rho u}{ds} 
- (\gamma v(u+vt^-)- 2E_c t^-) \frac{d\rho v}{ds} +\gamo (u+v t^-)\frac{d\rho E}{ds}  \right) ds \nonumber,
\eea
}
\end{linenomath}
with $s$ the curvilinear abscissa along the characteristic curve. The subparts of $\Xi C+ $ and $\Xi C-$
are also calculated to avoid any error
 in scale when discussing close to zero numerical values. For $\Xi C^+$, for example,
\begin{linenomath}
{\footnotesize
$$ \Xi C^+_1 = \int_{{\cal C}^+}  \left( \gamo E_c (u+vt^+) \frac{d\rho}{ds} \right) ds ~\qquad
\Xi C^+_2 = -\int_{{\cal C}^+} \left( (\gamma u(u+vt^+)-2 E_c ) \frac{d\rho u}{ds} \right) ds  $$
$$ \Xi C^+_3 = -\int_{{\cal C}^+}  \left( (\gamma v(u+vt^+)- 2E_c t^+) \frac{d\rho v}{ds} \right) ds ~\qquad
 \Xi C^+_4 = \int_{{\cal C}^+} \left( \gamo (u+v t^+)  \frac{d\rho E}{ds} \right) ds.  $$
}
\end{linenomath}
Of course, the $\Xi C^+ = \Xi C^+_1 +\Xi C^+_2+ \Xi C^+_3+\Xi C^+_4 $
(resp. $\Xi C^-= \Xi C^-_1 +\Xi C^-_2+ \Xi C^-_3+\Xi C^-_4 $) sum
is expected to be much smaller
than its subparts. The integration is performed in the information propagation direction (increasing $x$)
along the selected \cp and the selected \cm  presented in figure \ref{fig:geoflow}. 
Almost null values of $\Xi C^+$ and $\Xi C^-$ are indeed observed -- see figure \ref{fig:dCpmflow}.  
Equivalent successful verifications have been performed to assess the validity of (\ref{e_p2}) and (\ref{e_p3}).
They are not presented here for the sake of brevety. These two streamtraces (DCE) are equivalent to $dS=0$ and $dH=0$
 as discussed in \S 3.3.
 As a complementary verification, the relative variation of $H$ (all over the streamtrace)
 and $S$ (up to the first shockwave) are calculated. Relative errors of $3e^{-6}$ and $4e^{-6}$ are found.
%
%
 \subsection{Assessment of the (ACE) for the thrust-adjoint}
\begin{figure}[]
\begin{center}
\begin{minipage}{0.49\linewidth}
  \begin{center}
  \includegraphics[width=0.9\linewidth]{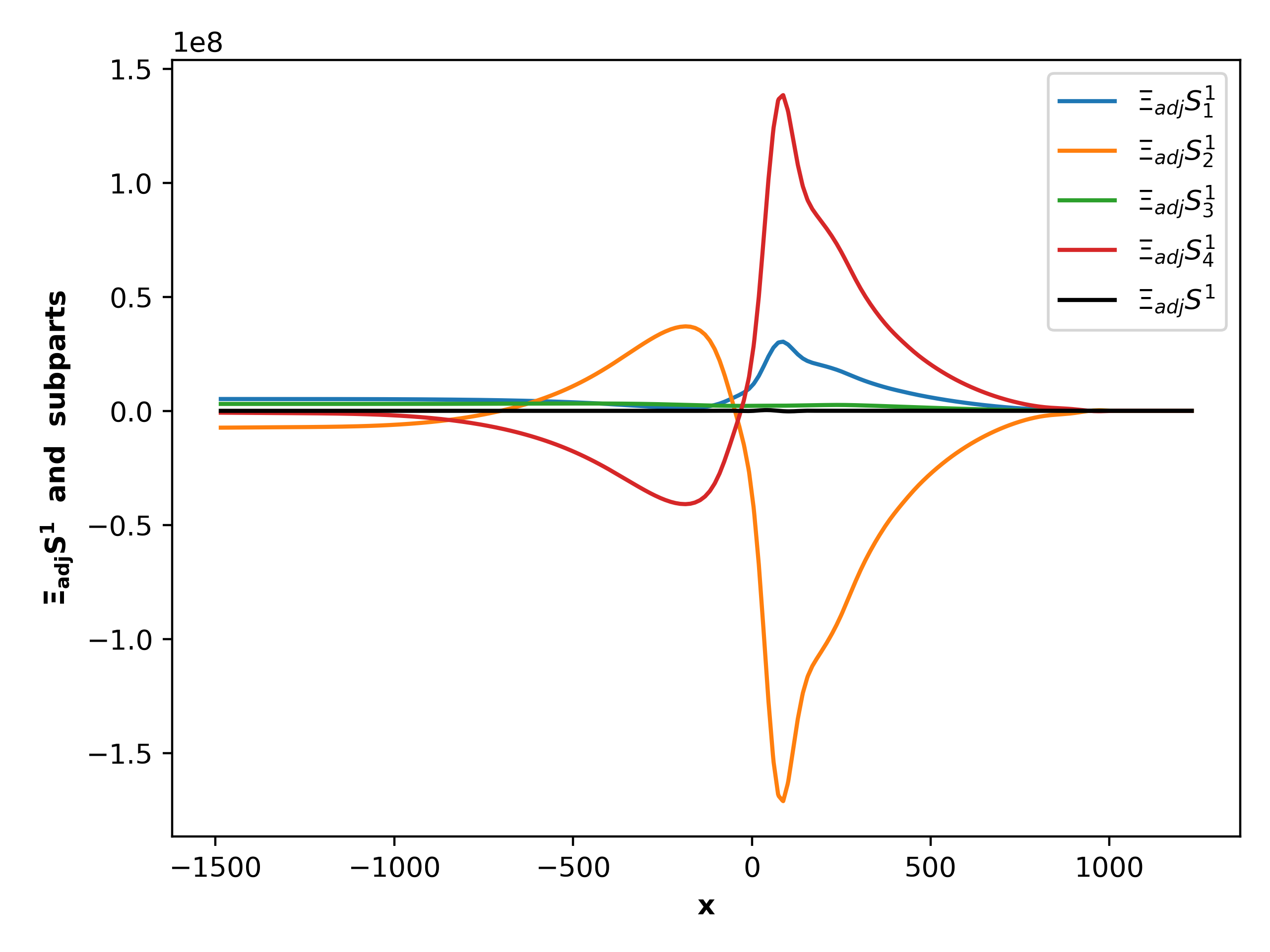}
  \end{center}
\end{minipage}%
\begin{minipage}{0.5\linewidth}
  \begin{center}
  \includegraphics[width=0.9 \linewidth]{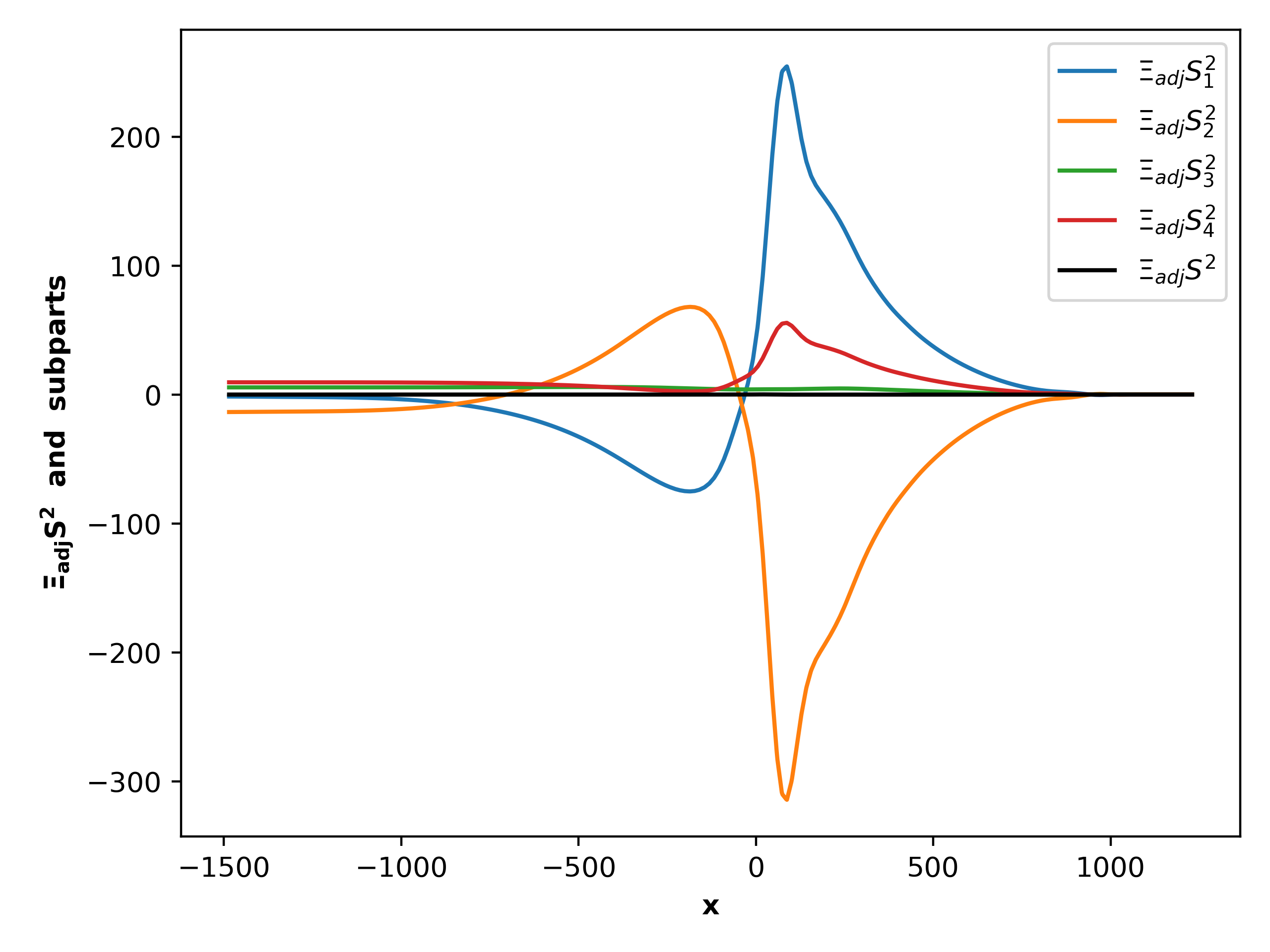}
  \end{center}
\end{minipage}
\begin{minipage}{0.49\linewidth}
  \begin{center}
  \includegraphics[width=0.9\linewidth]{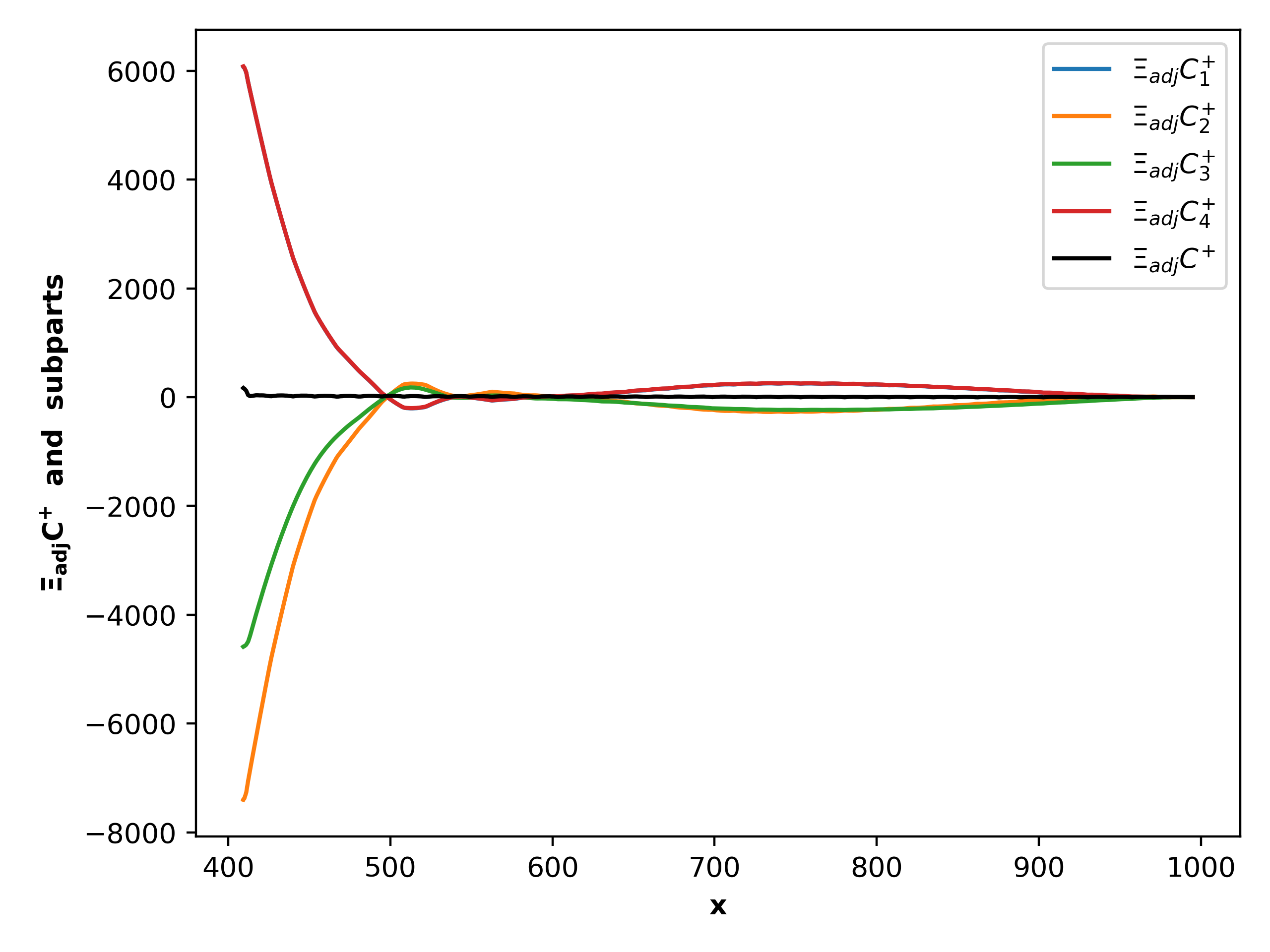}
  \end{center}
\end{minipage}%
\begin{minipage}{0.5\linewidth}
  \begin{center}
  \includegraphics[width=0.9 \linewidth]{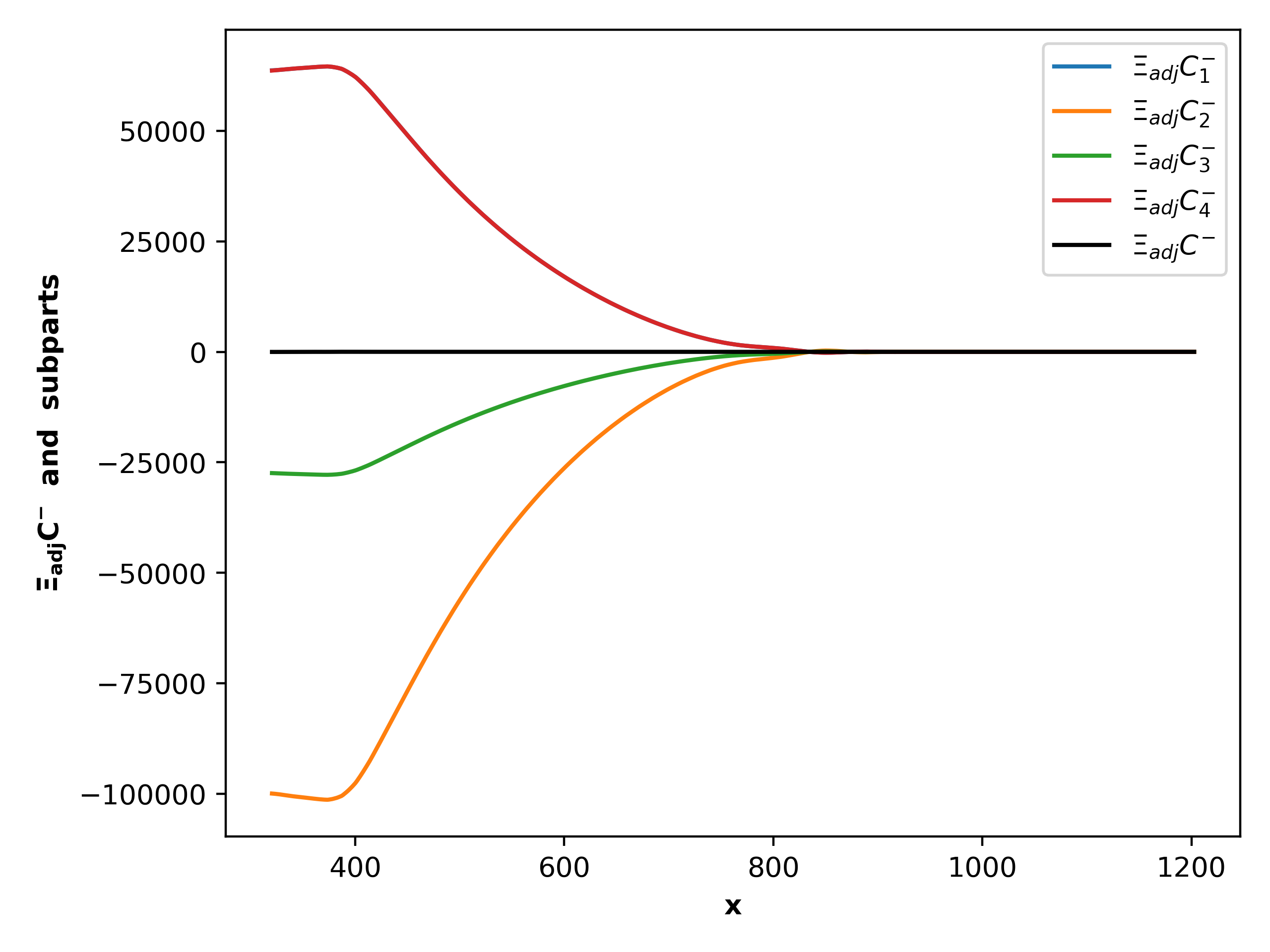}
  \end{center}
\end{minipage}
\caption{ Numerical assessment of the (ACE)\cite{JPDes_22}.
  $\Xi_{adj}S^1$ (upper left), $\Xi_{adj}S^2$ (upper right), $\Xi_{adj}C^+$ (bottom left), $\Xi_{adj}C^-$ (bottom right)
  and their subparts integrated along the selected \st, \cp and \cm. Method of verification: the black curve
should ideally coincide with the $x$ axis.}
\label{fig:dCadj}
\end{center}
\end{figure}

 For the thrust-adjoint, the assessment method is the same except that the (ACE) are integrated in the forward sense
 for the adjoint, that is, from the support of the functional output backwards w.r.t. the
 direction of the flow information propagation.
 The integrals to evaluate (as well as their subparts) from the discrete flow and thrust-adjoint
 fields are\cite{JPDes_22}:
{\footnotesize 
\bea
 &\Xi_{adj}S^1& = \int_{\cal S} \left(   E_c ~\frac{d\psi_1}{ds} +   H ( u~ \frac{d\psi_2}{ds} + v~ \frac{d\psi_3}{ds})  +   ~H^2  \frac{d \psi_4}{ds}  \right) ds    \\
 &\Xi_{adj}S^2& = \int_{\cal S} \left( \frac{d\psi_1}{ds} +    u~ \frac{d\psi_2}{ds} + v~ \frac{d\psi_3}{ds}  +   ~E_c  \frac{d \psi_4}{ds} \right)  ds  \\
 &\Xi_{adj}C^+& = \int_{{\cal C}^+}  \left(  (u+v t^+)\frac{d\psi_1}{ds} +  (u^2 + v^2) (\frac{d\psi_2}{ds} + t^+ \frac{d\psi_3}{ds} ) + H  (u+v t^+)\frac{d\psi_4}{ds}  \right) ds\\
&\Xi_{adj}C^-& = \int_{{\cal C}^-}  \left(  (u+v t^-)\frac{d\psi_1}{ds} +  (u^2 + v^2) (\frac{d\psi_2}{ds} + t^- \frac{d\psi_3}{ds} ) + H (u+v t^-)\frac{d\psi_4}{ds} \right) ds.
\eea
}
As in \S 4.2, the (CE) are numerically assessed by checking that $\Xi_{adj}S^1$,  $\Xi_{adj}S^2$, $\Xi_{adj}C^+$ and
$\Xi_{adj}C^-$
are much smaller than their subterms all along the characteristic curves. This property is actually well satisfied as can be seen
 in figure \ref{fig:dCadj}.

%
\section{Conclusion}
A few years ago, several authors had the intuition that the lift- and drag-adjoint fields of the supersonic areas of
2D inviscid flows were linked to the characteristic curves \cite{SarMetSip_15,Loz_17e}. This was recently confirmed
by the extension of the theory of characteristics to the continuous adjoint equations of 2D Euler flows \cite{JPDes_22}.
The application section of this paper provides a second numerical assessment of these equations.
\\
Besides, the derivation of the (ACE) in  \cite{JPDes_22} required the analytical calculation
of the minors of a $8\times8$ Cauchy problem matrix. The counterpart demonstration had never been applied for the search
of the (CE) of the flow (the analysis of a set of physical equations, in a frame attached to the local velocity, 
being the classical way to proceed). It appears that the search of the (DCE) in conservative variables
in the original frame of reference is feasible and not too complex.  The corresponding derivation is greatly
simplified by algebraic properties, presented in this contribution, which are
shared by the linear systems of the Cauchy problem
 involved in both, the (DCE) and (ACE) search.
%
%
\section*{Supplementary materials}
Three Python files checking the formulas of $K^l_{mx}$ , $K^l_{my}$, $\bar{K}^l_{mx}$,  $\bar{K}^l_{my}$. \\
Two Maple files checking the rank of the differential forms along \st ~ and \cp curves.
\section*{Funding}
The nozzle considered in \S 4 was designed in the framework of the SENECA project that is funded by the European Union’s
Horizon 2020 research and innovation programme under grant agreement No. 101006742,
project SENECA ((LTO) Noise and Emissions of Supersonic Aircraft).\\
This work was partially supported by the SONICE project, granted by the French Directorate General for Civil Aviation (DGAC).
\section*{Acknowledgements} 
elsA V5.0 used in \S 4 is ONERA-Safran property.\\
The authors warmly thank S\'ebastien Heib, Sana Amri, S\'ebastien Bourasseau,Thomas Hennion and Antoine Dumont for many fruitful discussions.
\section*{Authors contribution}
Investigation, K.Ancourt, J.Peter ; Software, J.Peter ; Validation, K.Ancourt, O.Atinault ; Writing, J.Peter, K.Ancourt, O.Atinault ; Formal Analysis, K.Ancourt.
%


\appendix
\section*{Appendix A}
%
 $\bullet$ The   $K^l_{2x}$ coefficients are expressed below
\begin{linenomath}
\beas
K^1_{2x}&=& -dx^3~\gamo ~t~E_c~\ww~( \gamo ~H -(\gamma+1)E_c)                                                 \\
K^2_{2x}&=& dx^3  \ww  (\gamo^2  t  u  E_c + \gamo E_c \ww + v^2  \ww + \gamo H v  - \gamo^2 ~H  t  u + \gamo  t  u  E_c )\\     
K^3_{2x}&=& -dx^3~t~\ww~(-u^2~\ww + \gamo~((\gamma+1)~v-t~u)~E_c + \gamo ~H~(t~u - \gamo~v)) \\     
K^4_{2x}&=& -dx^3~\gamo~t~\ww~( -\gamo ~H+(\gamma+1)E_c)                                                     
\eeas
\end{linenomath}
$\bullet$ The $K^l_{3x}$  coefficients are expressed below 
\begin{linenomath}
\beas
K^1_{3x}&=&  -dx^3~\gamo~t^2~E_c~\ww~(-\gamo ~H + (\gamma+1)E_c) \\                                                     
K^2_{3x}&=&  -dx^3~t~\ww~( \gamo~((\gamma+1)~t~u-v)~E_c + v^2~\ww + \gamo ~H~(v - \gamo~t~u) )\\      
K^3_{3x}&=& dx ^3~\ww~(-v^3 + 3~t~u~v^2 -\gamo ~H~\ww - 2~t^2~u^2~v+\gamo^2~t^2~v~E_c +\gamo ~H~t^2~v \\
&-&\gamo^2 ~H~t^2~v + \gamo~t^2~v~E_c+\gamo~E_c~\ww) \\      
K^4_{3x}&=&    -dx^3~\gamo ~t^2~\ww~(\gamo ~H- (\gamma+1)~E_c)
\eeas
\end{linenomath}
$\bullet$  The $K^l_{4x}$ coefficients are expressed below
\begin{linenomath}
\beas
K^1_{4x}&=&  dx^3~ \gamo~t ~E_c~ ~ \ww ~ (u+t~v)~(\gamo~E_c+(2-\gamma)~H) \\                                                      
 K_{4x}^2 &=& dx^3 ~t~ \ww~(2~\gamo ~E_c + \gamma~\ww~v)~(\gamo~E_c + (2-\gamma)H) \\
 K_{4x}^3 &=& -dx^3~t~\ww~(-u~\ww + \gamo~t~v^2 + \gamo~u~v)~(\gamo~E_c +(2-\gamma)H) \\    
K^4_{4x}&=& dx^3\ww(-\gamo H \ww - \gamo v E_c - v\ww^2 - \gamo^2 t u E_c - \gamo t^2 v E_c \\
&-& \gamo^2 t^2 v E_c +  \gamo^2 Ht u +  \gamo^2 H t^2 v  )            
\eeas
\end{linenomath}
$\bullet$ The $K^l_{1y}$  coefficients are expressed below 
\begin{linenomath}
\beas
K^1_{1y}&=& dx^3~\ww~(u~\ww^2 + 2 \gamo~u~E_c + \gamo~t~v~E_c + \gamo~t^2~u~E_c - \gamo ~H~(1+t^2)~u )  \\                                                      
K^2_{1y}&=& dx^3~\ww~(\gamma ~ \ww~v + 2~\gamo ~E_c) \\  
K^3_{1y}&=& dx^3~\ww~(-\gamo~v~(t~v+u)+ \ww~u) \\ 
K^4_{1y}&=&  -dx^3~\gamo~\ww~(t~v + u)                                                     
\eeas
\end{linenomath}
$\bullet$ The $K^l_{2y}$  coefficients are expressed below
\begin{linenomath}
\beas
K^1_{2y}&=& dx^3~\gamo~E_c~\ww~( \gamo ~ H - (\gamma+1)~E_c) \\ 
K^2_{2y}&=& dx^3~\ww(\gamo ~E_c~(t\ww -\gamma~u) +t^2~u^3  + 2uv^2 - 3~t~u^2~v + \gamo ~H~((\gamma-2)u-t\ww) ) \\
K^3_{2y}&=& dx^3~\ww~(-u^2~\ww + \gamo~((\gamma+1)~v-t~u)~E_c + \gamo ~H~(t~u - \gamo~v)) \\  
K^4_{2y}&=& dx^3~\gamo~\ww~( -\gamo ~H+(\gamma+1)~E_c) 
\eeas
\end{linenomath}
$\bullet$ The  $K^l_{3y}$ coefficients are expressed below 
\begin{linenomath}  
\beas
 K_{3y}^1 &=& dx^3~\gamo~E_c~\ww~t~(-\gamo ~H + (\gamma+1)~ E_c)\\
 K_{3y}^2 &=& dx^3~\ww~( \gamo~((\gamma+1)~t~u-v)~E_c + v^2~\ww + \gamo ~H~(v - \gamo~t~u) ) \\
 K_{3y}^3 &=& dx^3~\ww~t~(\gamo ~H~(\gamo~v-t~u) - 2~\gamo~v~E_c - \gamo^2~v~E_c + u^2~\ww + \gamo~t~u~E_c) \\
 K_{3y}^4 &=&  dx^3~\gamo~\ww~t~(\gamo ~H - (\gamma+1)~E_c)
 \eeas
 \end{linenomath}
$\bullet $ The  $K^l_{4y}$ coefficients  are expressed below
\begin{linenomath} 
\beas
 K_{4y}^1 &=& - dx^3~ \gamo ~E_c~ ~ \ww ~ (u+t~v)~(\gamo~E_c+(2-\gamma)~H) \\
 K_{4y}^2 &=& -dx^3 ~ \ww~(2~\gamo ~E_c + \gamma~\ww~v)~(\gamo~E_c + (2-\gamma)~H) \\
 K_{4y}^3 &=& dx^3~\ww~(-u~\ww + \gamo~t~v^2 + \gamo~u~v)~(\gamo~E_c +(2-\gamma)~H) \\
 K_{4y}^4 &=& dx^3 ~ \ww ~ (u~\ww^2 + \gamma~\gamo~u~E_c + \gamo~t^2~u~E_c + \gamo^2~t~v~E_c -\gamo^2 ~H~u - \gamo^2 ~H~t~v \\
 &-& \gamo ~H~t^2~u + \gamo~H~t~v )  
 \eeas
 \end{linenomath}
%
%
\section*{Appendix B}

Let us prove the equivalence between (\ref{ebase}) and  (\ref{e_dcpcm}) in case of a  \cp. Equation (\ref{ebase}) reads
\begin{linenomath}
$$  d \phi - \sqrt{M^2-1} \frac{d\normU}{\normU} - \sqrt{M^2-1} \frac{TdS - dH}{\normU^2}  = 0, $$  
\end{linenomath}
with 
\begin{linenomath}
  $$ \phi = \arctan(\frac{\rho v}{\rho u}) \textrm{~~~~~~~and~~~~~~~}~~~ T= \frac{p}{\rho r}
  \textrm{~~~so~that~~~}TdS = \frac{p}{\rho \gamo}d(ln(\frac{p}{\rho^{\gamma}}))= \frac{p}{\rho \gamo}(\frac{dp}{p}- \gamma \frac{d\rho}{\rho}). $$
\end{linenomath}
Moving from differentiation w.r.t. primitive variables to  differentiation w.r.t. conservative variables yields  
\begin{linenomath}
\beq
\gamo E_c \sqrt{M^2-1} ~ d \rho -(v+ \gamo u \sqrt{M^2-1}) ~d \rho u + (u - \gamo v \sqrt{M^2-1} ) ~d \rho v + \gamo \sqrt{M^2-1} ~d \rho E =0 .
\eeq
\end{linenomath}
Besides, equation (\ref{e_dcpcm}) reads
\begin{linenomath}
$$\gamo  E_c (u+t^{+}v) ~d\rho -(\gamma u (u+t^{+}v)- 2 E_c) ~d\rho u -(\gamma v (u+t^{+} v)- 2 E_c t^{+}) ~d\rho v+ \gamo(u+t^{+} v) ~d\rho E = 0 $$
\end{linenomath}
It appears that  the coefficient of the last two equations are proportional:
\begin{linenomath}
\beq
 \frac{\gamo  E_c (u+t^{+}v)}{\gamo E_c \sqrt{M^2-1} }=\frac{-(\gamma u (u+t^{+}v)- 2 E_c)}{-(v+ \gamo u \sqrt{M^2-1} )}=\frac{-(\gamma v (u+t^{+} v)- 2 E_c t^{+})}{(u - \gamo v \sqrt{M^2-1} )}=\frac{\gamo(u+t^{+} v)}{\gamo \sqrt{M^2-1}}=(u+t^+v) ~\tan{\beta}.
\eeq 
\end{linenomath}

\normalsize
\section*{Biblography}
\bibliography{my_biblio}{}
\bibliographystyle{nature}

\end{document}